\documentclass[twocolumn,showpacs,preprintnumbers,amsmath,amssymb,superscriptaddress,prl]{revtex4}

\usepackage{graphicx}
\usepackage{dcolumn}
\usepackage{bm}
\usepackage{txfonts, color}

\newcommand{\rmd}{{\rm d}}
\newcommand{\rme}{{\rm e}}

\usepackage{amsmath}	
\begin{document}


\title{Dynamic response of a mesoscopic capacitor in the presence of strong electron interactions}

\author{Yuji Hamamoto}
\email{hamamoto@issp.u-tokyo.ac.jp}
 \affiliation{%
Institute for Solid State Physics, University of Tokyo, Kashiwa, Chiba 277-8581, Japan%
}%
\author{Thibaut Jonckheere}%
 \affiliation{%
Centre de Physique Th\'eorique, Case 907 Luminy, 13288 Marseille cedex 9, France%
}%

\author{Takeo Kato}
 \affiliation{%
Institute for Solid State Physics, University of Tokyo, Kashiwa, Chiba 277-8581, Japan%
}%

\author{Thierry Martin}
 \affiliation{%
Centre de Physique Th\'eorique, Case 907 Luminy, 13288 Marseille cedex 9, France%
}%
\affiliation{%
Universit\'e de la M\'edit\'erann\'ee, 13288 Marseille cedex 9, France
}%
\date{\today}

\begin{abstract}
We consider a one dimensional mesoscopic capacitor
in the presence of strong electron interactions
and compute its admittance in order to probe the universal nature of the 
relaxation resistance. We use a combination of perturbation theory,
renormalization group arguments,
and quantum Monte Carlo calculation to treat the whole parameter range
of dot-lead coupling.
The relaxation resistance is universal even in the presence of strong
Coulomb blockade
when the interactions in the wire are sufficiently weak.
We predict and observe a quantum phase transition to an incoherent regime
for a Luttinger parameter $K<1/2$.
Results could be tested using a quantum dot coupled to an edge state in the
fractional quantum Hall effect.    
\end{abstract}


\maketitle


The dynamical response of mesoscopic conductors 
constitutes a mostly unexplored area of coherent quantum transport, 
which has recently led to groundbreaking 
experiments~\cite{gabelli}.
The mesoscopic capacitor~\cite{buttiker_pretre} is one of its
elementary building blocks: a quantum dot 
influenced by an AC gate voltage, which is put in contact with an 
electron reservoir. It has been studied so far at the single 
electron level, with possible mean field generalizations~\cite{nigg}.
Both the capacitance $C_\mu$ and the relaxation resistance $R_q$,  
obtained from the low frequency expansion 
of the admittance $G(\omega)\approx -i\omega C_\mu +\omega^2C_\mu^2 R_q$,  
are fundamentally affected by the quantum coherence of the device.  
At zero temperature, a single spin polarized channel yields 
a relaxation resistance $R_q=h/(2e^2)$, which is independent of the 
dot-reservoir connection. 
Ref.~\cite{gabelli} has confirmed
this result for a quantum dot with 
weak charging energy.

However, quantum dots with reduced size exhibit strong Coulomb blockade,
and there is also a clear need to analyze whether electron-electron 
interactions in the lead are relevant.
Here, taking rigorously these aspects into account,
we prove that there is quantum phase transition
from a coherent to an incoherent regime, where a relaxation 
resistance cannot be defined. 
For weak interactions, the universal behavior is recovered even 
in the presence of strong Coulomb blockade.

\begin{figure}[b]
 \includegraphics[width=.8\columnwidth]{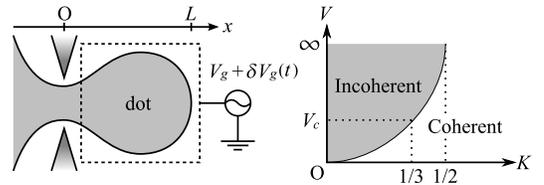}
 \caption{\label{fig:dot} 
Left: schematic view of the mesoscopic capacitor: a 1D quantum dot, capacitively coupled
to a gate with time-dependent voltage $V_g + \delta V_g(t)$.
Right: schematic phase diagram in the degenerate case.
$V_c$ denotes the critical backscattering strength for $K=1/3$.}
\end{figure}

We consider a quantum dot (Fig.~\ref{fig:dot}) connected to a reservoir 
modelled by a Luttinger liquid lead, which allows to account exactly 
for Coulomb blockade effects. We discuss separately the absence 
(Luttinger parameter $K=1$) or the presence ($K<1$) of interaction 
in the adjacent lead. This setup and the underlying physics is 
similar to that studied in Ref.~\cite{Furusaki02}, where attention 
was solely focused on the static occupation of a resonant level. 
Here we show that below $K=1/2$ the Kosterlitz Thouless type phase 
transition driven by the dot-lead tunneling strength triggers a 
transition of dynamical transport from a coherent to an incoherent 
regime, hence provoking a deviation 
from the universal $R_q=h/(2e^2)$. We use a combination of analytical 
(perturbation theory, renormalization group) and numerical (quantum Monte Carlo) 
approaches  to monitor the capacitance and the relaxation resistance over the whole 
range of dot-lead connection. The present results can be applied 
to carbon nanotube quantum wires as well as dots defined in 
the fractional quantum Hall effect (FQHE).

The starting point is the Hamiltonian for a non-chiral, semi-infinite
Luttinger liquid~\cite{kane_fisher_92} where the dot region
corresponds to the interval $[0,L]$:
\begin{align}
H=&\int_{-\infty}^L\frac{\rmd x}{2\pi}\left[\frac{v_F}{K^2}
\left(\frac{\partial\phi}{\partial x}\right)^2
+v_F \left(\frac{\partial\theta}{\partial x}\right)^2\right]
-V\cos[2\phi(x=0)]
\nonumber \\
 &+\frac{1}{\pi^2}E_C\left[\phi(x=0)
-\left(\frac{\pi CV_g}{|e|}+k_FL\right)\right]^2 .
\end{align}
The first part is the kinetic part, followed by the backscattering 
term at $x=0$ (strength $V$), 
and finally the contribution from the charging energy with $E_C\equiv e^2/(2 C)$
($C$ is the geometrical capacitance).
The canonically conjugated fields $\phi$ and $\theta$ satisfy
the commutation relation $[\phi(x),\theta(x')]=(i\pi/2)\mbox{sgn}(x-x')$.
$V$ is the backscattering strength on the point contact.
$E_C\equiv e^2/(2 C)$ denotes the charging energy
The time dependent gate voltage oscillates around $V_g$.
Using the Matsubara imaginary time path integral formulation, the quadratic 
degrees of freedom away from $x=0$ can be integrated out.
The kinetic part of the effective action then reads
$S_{kin}=(\pi K\beta)^{-1}\sum_{\omega_n}
|\omega_n|/(1-\rme^{-2\pi K|\omega_n|/\Delta})
|\tilde\phi(\omega_n)|^2$,
where $\tilde{\phi}(\omega_n)$ is the Fourier transform of 
$\phi(\tau)$ (now specified at $x=0$), and $\Delta\equiv\pi v_F/L$
is the level spacing.
The same action can be derived alternatively starting from a 
single chiral Luttinger liquid ``loop'', hence the relevance for
the FQHE regime~\cite{FQHE_explain}.
Within linear response in the oscillating gate voltage,
the (imaginary frequency) admittance can be expressed as: 
\begin{gather}
 G(i\omega_n)=\frac{e^2}{h}\frac{2 |\omega_n|}{\pi}
\int_0^\beta\rmd\tau\langle\phi(\tau)\phi(0)\rangle\rme^{i\omega_n\tau}.
\label{eq:admittance}
\end{gather}
The dynamical conductance is obtained by analytic continuation
$G(\omega) = G(i\omega_n \rightarrow \omega + i\delta$), 
while the capacitance reads:
\begin{gather}
C_\mu=\frac{e^2}{\pi^2}\beta[\langle\bar\phi^2\rangle
-\langle\bar\phi\rangle^2]\qquad
\left(\bar\phi\equiv\frac{1}{\beta}\int_0^\beta\rmd\tau\phi(\tau)\right)
.\label{eq:capacitance}
\end{gather}

We start with a discussion of the weak barrier case, using perturbation theory
in $V/D$ (bandwidth $D$). The capacitance and relaxation resistance are
derived as an expansion in orders of $V$, $C_\mu=C_{\mu}^{(0)}+C_{\mu}^{(1)}+C_{\mu}^{(2)}+...$ and 
$R_q=R_q^{(0)}+R_q^{(1)}+R_q^{(2)}+...$.
Introducing
\begin{gather}
a_n = \frac{1}{\pi K \beta}
\left( \frac{|\omega_n|}{1-\rme^{-2\pi K|\omega_n|/\Delta}} + \frac{E_C K}{\pi} \right)  , 
\end{gather}
one obtains to zeroth, first and second order:
\begin{align}
G^{(0)}(i \omega_n)&=\frac{e^2}{h} \frac{|\omega_n|}{\pi}
\frac{1}{\beta a_n} \label{eq:G0pert}\\
G^{(1)}(i\omega_n)&=-\frac{e^2}{h} \frac{|\omega_n|}{\pi} 2 V
   \sqrt{F_{+}(0)} \frac{1}{\beta^2 a_n^2} \cos(2 \pi N)\label{eq:G1pert}\\ 
G^{(2)}(i\omega_n)&=\frac{e^2}{h} \frac{|\omega_n|}{\pi} 2 V^2 
  F_{+}(0) \frac{1}{\beta^2 a_n^2}I(\omega_n), \label{eq:G2pert}
\end{align}
where we defined:
\begin{gather}
F_{\pm}(v) = \exp\left[ -\sum_n \frac{\pm 2}{\beta^2 a_n} \cos(\omega_n
 v) \right]\\
I(\omega_n)=
\int_0^{\beta/2}{\rm d}v\left[ \cos(4 \pi N) V_{+}(\omega_n,v) +  V_{-}(\omega_n,v)\right]\\
V_{\pm}(\omega_n,v) = 1- F_{\pm}(v) \left(1 \pm \cos(\omega_n v)\right)  \\
N=\frac{k_FL}{\pi}+\frac{|e|V_g}{2E_T}\quad
\mbox{with}\quad E_T=E_C+\frac{\Delta}{2K^2}.
\end{gather} 
The $\sum_n$ in $F_{\pm}$ is limited by $D$.
 From Eqs.~(\ref{eq:G0pert})-(\ref{eq:G2pert}), the capacitance at low temperature becomes:
\begin{align}
C^{(0)}_{\mu} &=\frac{e^2}{h} \frac{\pi}{E_T}
\label{eq:C0}\\
C^{(1)}_{\mu} &= -C^{(0)}_{\mu}  \frac{\pi^2}{E_T}
2V\sqrt{F_{+}(0)} \cos(2 \pi N) \\
C^{(2)}_{\mu} &= C^{(0)}_{\mu}  \frac{\pi^2 }{E_T}  
2V^2 F_{+}(0)I(\omega_n\rightarrow 0)\cos(4 \pi N)  .
\end{align}
It is clear from these expressions that the total capacitance $C_{\mu}$
is a periodic function of $N$, with period 1.
Below we focus on the interval $0\le N<1$.
The results for the relaxation resistance, at low temperature, are simple since the computation of the
first and second order contribution shows that they vanish:
  \begin{equation}
  R_q^{(0)} + R_q^{(1)} + R_q^{(2)} = R_q^{(0)}
= \frac{h}{2 e^2}   \frac{1}{K}.\label{eq:Rq0}
\end{equation}
The charging energy thus does not modify the value of relaxation resistance,
while electron interactions in the lead introduce a trivial factor $1/K$. 
At zero temperature, the sums and integration of
Eqs.~(\ref{eq:G0pert})-(\ref{eq:G2pert}) can be done
analytically in certain cases. For example,
when $E_C=0$ and $K=1$,
one has $F_{+}(0) = (\Delta/(2 \pi D) )^2$ ($D \gg \Delta$),
and one can show that the result for the capacitance is 
$C_\mu= (e^2/\Delta) (1 - 2 r \cos(2 \pi N) + 2 r^2 \cos(4 \pi N))$,
with $r= \pi V/D$.
This coincides with the  development of the non-interacting formula
found in Ref.~[\onlinecite{gabelli}]
in powers of the reflection coefficient $r$:
$C_\mu = (e^2/\Delta) (1-r^2)/(1 - 2 r \cos(2 \pi N) + r^2)$. In the more general case of non-zero $E_C$, and $K \neq 1$, one has
$F_{+}(0) \sim (E_T K/(\pi D))^{2K}$, and the integration of Eq.~(\ref{eq:G2pert}) has to be computed numerically.

\begin{figure}[t]
 \includegraphics[width=\columnwidth]{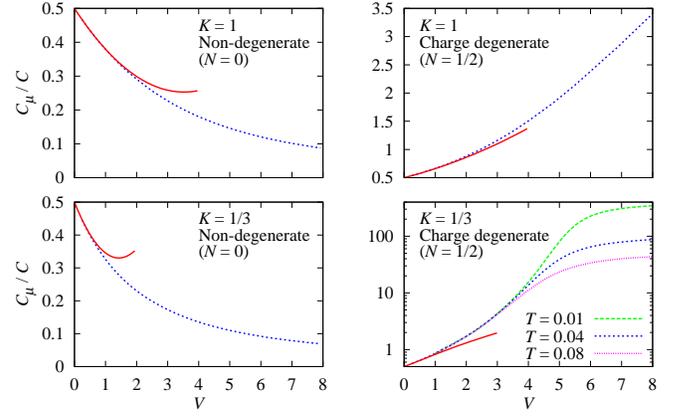}
 \caption{\label{fig:capacitance}(color online) Capacitance $C_{\mu}$
as a function of the backscattering strength $V$,
obtained at temperature $T=0.04$ with Monte Carlo computations (dashed lines).
The solid lines show the predictions of the perturbative calculations up to
second order.
We take $E_C/\pi^2$ as the unit of energy and
use the parameters $D\equiv2\pi J/\beta=8\pi$ and $\Delta/(2K^2)=\pi^2$.
Temperature dependence is shown
in the bottom right panel,
where the vertical axis is measured on a logarithmic scale.}
\end{figure}

The perturbation theory thus proves the universality of the charge
relaxation resistance
in the weak barrier limit
even in the presence of interactions.
To study the non-perturbative regime, the path-integral Monte Carlo
method is applied to
the action for the discretized path $\phi(\tau=j\beta/J)$
$(j = 0,1,\cdots, J-1)$.
We estimate thermal average by generating discretized paths
using local update in the Fourier space and
the cluster update~\cite{Werner05,Hamamoto}.
The top (bottom) row of Fig.~\ref{fig:capacitance} shows
the calculated capacitance $C_\mu$ as a function of $V$ at $T=0.04$ and
$K=1$ ($K=1/3$).
The left and right columns correspond to the non-degenerate case
($N=0$)
and the charge degenerate case ($N=1/2$),
respectively.
With increasing $V$ the Coulomb staircase becomes sharper,
which results in the decrease (increase) in
$C_\mu\equiv\partial\langle Q\rangle/\partial V_g$
in the case of $N=0$ ($N=1/2$).
The second-order perturbation theory,
shown as solid lines, displays an excellent
agreement for small $V$.
Especially, it is remarkable that only for the case of $K=1/3$ and $N=1/2$
(the right bottom panel of Fig.~\ref{fig:capacitance}),
$C_\mu$ exhibits an abrupt increase at a finite $V$,
signaling a possible transition.
One can see that $C_\mu$ grows as $\propto 1/T$
in the large barrier region.

To reveal the origin of the transition behavior,
we next examine the strong barrier limit
using an instanton method 
which was developed for the Kondo model~\cite{anderson}.
Near the degeneracy point $N=1/2$,
the configuration of the bosonic field $\phi$ can be represented
in the dilute instanton gas approximation 
\begin{equation}
\phi(\tau)\simeq \pi \sum_{j=1}^{2n} s_j \Theta(\tau-\tau_j)
+\frac{\pi}{2}(1-s),
\label{phi_instanton}
\end{equation}
where $s_j=s(-1)^{j-1}$, and $s\sim 1$ denotes the separation between the well minima
($\Theta$ is the step function). 
Inserting Eq. (\ref{phi_instanton}) in the full effective action, the partition
function becomes:  
\begin{gather}
\begin{split}
 Z&=\sum_{n=0}^\infty t^{2n}\int_0^\beta\rmd\tau_{2n}
\int_0^{\tau_{2n}}\rmd\tau_{2n-1}\cdots\int_0^{\tau_2}\rmd\tau_1\\
&\phantom{=}\times\exp\left[\frac{1}{2K}\sum_{j\ne k}s_js_k
\log\frac{|\tau_j-\tau_k|}{\tau_c}-u\sum_js_j\tau_j\right],
\end{split}
\end{gather}
where $t$ is the tunneling amplitude between the well minima,
and $\tau_c$ is the short-time cutoff.
$u=(2N-1)E_T$ denotes the deviation from the degeneracy point.
Note the similarity between this partition function
and that which was proposed
in the context of dissipative Josephson junctions~\cite{chakravarty}.
One can therefore identify the scaling equations
\footnote{We assume that the local scatterer at $x=0$
does not renormalize 
significantly the bulk interaction parameter $K$.}:
\begin{gather}
 \frac{\rmd t}{\rmd l}=\left(1-\frac{s^2}{2K}\right)t,\label{eq:scaling} \quad
\frac{\rmd s^2}{\rmd l}=-4s^2t^2,\\
\frac{\rmd us}{\rmd l}=us(1-2t^2)\label{eq:third}
\end{gather}
which are familiar in the context of a Kosterlitz Thouless transition
in the two-dimensional XY model.
No further arguments are needed when one deviates from the degeneracy
point:
since $t$ is small, starting from $u\neq 0$,
Eq.~(\ref{eq:third})
predicts that $u$ will further increase, leading the system
further away from the degeneracy point. This means that $\phi$ will be
trapped in an effective harmonic potential, and one thus recover the
result of Eq.~(\ref{eq:G0pert}), which is therefore universal.
For the charge degenerate case $N=1/2$,
the transition corresponds to a Kondo type transition
associated with the charge pseudo spin on the dot.
Eqs.~(\ref{eq:scaling}) determine the tendency 
of the dot-lead transmission as temperature is lowered;
$(t,s^2)$ flows along one of the hyperbolic curves
$B^2-4t^2={\rm const}.$, where $B\equiv 1-s^2/(2K)$.
For $K>1/2$, the tunneling strength always grows upon reducing 
the temperature, 
and the system reaches the Kondo fixed point where the dot is strongly 
coupled to the reservoir.
An electron freely tunnels in and out of the dot 
irrespective of the initial tunneling strength. In particular at 
$K=1$ this implies that the charge relaxation resistance
is universal, i.e., $R_q=h/(2e^2)$, 
as a consequence of the unitary limit of the underlying Kondo model. 
On the other hand for $K<1/2$, there is the possibility that at a critical,
sufficiently weak transmission $t$ (``large'' $V$),
the RG flow always drives the system into 
a weak coupling configuration with specified charge.
Then the charge fluctuation remains finite, i.e.,
$\langle\bar\phi^2\rangle-\langle\bar\phi\rangle^2\simeq(\pi s/2)^2$,
so that the capacitance diverges as $\propto 1/T$ at low temperatures
[see Eq.~(\ref{eq:capacitance})].
This explains the transition observed for the capacitance
in the strongly interacting case.

\begin{figure}[t]
 \includegraphics[width=.8\columnwidth]{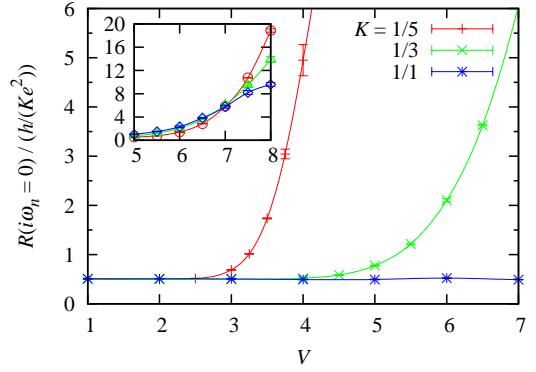}
\caption{\label{fig:resistance}(color online)
Extrapolated value $R(i\omega_n\rightarrow 0)$
as a function of $V$ at temperature $T=0.04$
in the degenerate case.
For $D$ and $\Delta/(2K^2)$,
we use the same parameters as in Fig.~\ref{fig:capacitance}.
Inset: $R(i\omega_n\rightarrow 0)$
for $K=1/3$ at $T=0.01(\circ)$, $0.04(\triangle)$ and $0.08(\diamond)$
in the vicinity of the location where the crossing occurs.}
\end{figure}

We now describe the effect of the KT transition 
on the dynamical properties.
If $\omega\ll1/\tau_{RC}$ holds (with $RC$ time $\tau_{RC}$),
the charge relaxation
resistance can be defined in the low-frequency expansion
$G(\omega)=-i\omega C_\mu+\omega^2C_\mu^2R_q+\mathcal{O}(\omega^3)$.
However, the validity of this expansion is not obvious,
since the KT transition may influence $\tau_{RC}$ itself.
Instead, we investigate the low-frequency resistance
using
\begin{gather}
 R(i\omega_n)\equiv\frac{1}{G(i\omega_n)}-\frac{1}{\omega_nC_\mu},
\end{gather}
where $G(i\omega_n)$ and $C_\mu$ are defined in Eqs.~(\ref{eq:admittance})
and (\ref{eq:capacitance}), respectively.
The extrapolation $R(i\omega_n\rightarrow0)$
gives the real part of the impedance
in the low-frequency limit, hence $\tau_{RC}=R(0)C_\mu$.
In Fig.~\ref{fig:resistance}, we plot $R(0)$ for $K=1, 1/3$ and $1/5$
as a function of $V$ at temperature $T=0.04$.
For $K=1$, $R(0)$ equals $h/(2e^2)$ irrespective of $V$,
in agreement with the universal charge relaxation
resistance~\cite{buttiker_pretre,nigg}.
For $K=1/3$ and $1/5$, the universality is observed
in the weak barrier region,
whereas $R(0)$ is abruptly enhanced with increasing $V$,
reflecting the RG flow to the weak coupling regime
due to the KT transition.
The temperature dependence of $R(0)$ for $K=1/3$
is shown in the inset of Fig~\ref{fig:resistance},
which indicates that $R(0)$ diverges as $T\rightarrow0$
in the strong barrier region.

\begin{figure}[t]
 \includegraphics[width=.8\columnwidth]{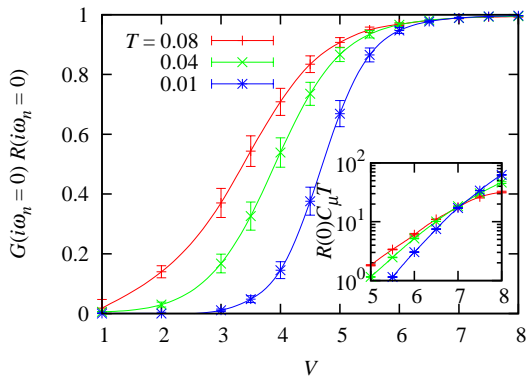}
\caption{\label{fig:GR}(color online)
Product $G(i\omega_n=0) R(i\omega_n=0)$ for $K=1/3$
as a function of the backscattering strength $V$, for
different temperatures $T$.
For $D$ and $\Delta/(2K^2)$,
we use the same parameters as in Fig.~\ref{fig:capacitance}.
Inset: $R(0) C_{\mu} T$ as a function of $V$, for different $T$; the crossing
of the curves gives a good estimate of the transition point.}
\end{figure}

The KT transition plays a crucial role
in the relevance of the universal charge relaxation resistance.
If $2t+B>0$,
the system scales to the weak barrier limit,
where $\tau_{RC}$ is independent of temperature.
If $2t+B<0$, on the other hand,
the scaling equations (\ref{eq:scaling}) predict
$s^2\rightarrow{\rm const.}$ and $t\propto T^{-B}$,
so that $\tau_{RC}$ roughly scales as
$\propto T^{-1}(T^{2B}+{\rm const.})$,
which grows faster than the (thermal) coherence time
$\tau_{\rm coh}\propto1/T$ as temperature is lowered.
These observations suggest that
if $2t+B>0$ coherent transport
can be realized by lowering temperature to guarantee
$\tau_{RC}<\tau_{\rm coh}$,
while if $2t+B<0$
electronic transport in the dot decoheres
before charge relaxation is achieved.
In the latter case, the quantum dot effectively acts as a reservoir
and consequently the dynamical property of the system
is governed by transport through the point contact
between the two ``reservoirs''.
Therefore the $V$-dependent low-frequency resistance observed
in the inset of Fig.~\ref{fig:resistance} reflects
the revival of the Landauer-type transport.
To see this behavior more clearly,
we plot in Fig.~\ref{fig:GR}
the product $G(i\omega_n\rightarrow 0)R(i\omega_n\rightarrow 0)$
for $K=1/3$ as a function of $V$.
In the strong barrier region,
$G(0)$ is finite and equal to $[R(0)]^{-1}$,
which is a familiar property of transport through a point contact.
Upon decreasing $V$, however, $G(0)R(0)$ is suppressed
since $G(0)$ decays to zero because of charging up,
although $R(0)\rightarrow R_q$ is finite.
Moreover, we see that the coherent region $G(0)R(0)=0$
extends to larger $V$ upon lowering temperature.
Finally, we determine the phase boundary of the coherent-incoherent
transition by tracing the temperature dependence of the ratio
$\tau_{RC}/\tau_{\rm coh}=R(0)C_\mu T$.
The above discussion suggests that there exists
a critical backscattering strength $V_c$,
below which $\tau_{RC}/\tau_{\rm coh}$ decays to zero,
while it diverges otherwise
(see the right panel in Fig.~\ref{fig:dot}).
From the inset of Fig.~\ref{fig:GR}, the critical value
is estimated as $V_c\simeq 7$.



In conclusion, the study of the mesoscopic capacitor
in the presence of strong electron electron interaction shows
that the relaxation resistance for a dot connected to Luttinger liquid is 
universal $R_q\equiv h/(2e^2K)$
as long as interactions are sufficiently weak.
Below $K<1/2$, this resistance is governed
by the strength of the dot-lead coupling:
at the charge degeneracy point,
there is a critical coupling strength, governed by a KT type phase transition, 
below which the dot acts as an incoherent reservoir and 
the low-frequency resistance exceeds
the universal value. 
In this incoherent regime,
the charge relaxation resistance cannot be defined anymore 
due to the divergence of the $RC$ time. 

Results could be probed experimentally using quantum dots connected 
to an edge state in the FQHE regime. Another experimental probe could use
one dimensional quantum wires (non chiral Luttinger liquids)
with the limitation that the operating frequency would have to be larger
than the inverse time of flight within the wire,
in order to avoid renormalization effects due to eventual Fermi 
liquid leads connected to this wire~\cite{maslov_stone}.     

Y.H. and T.K. are grateful to T. Fujii for valuable discussions.
Y.H. acknowledges the support of the Japan Society for the Promotion
of Science.
This research was partially supported by JSPS and MAE under the
Japan-France
Integrated Action Program (SAKURA) and by Grant-in-Aid for Young
Scientists (B)
(No. 21740220) from the Ministry of Education, Science, Sports and
Culture. It was also supported by ANR-PNANO Contract MolSpinTronics,
No. ANR-06-NANO-27.
The computation in this work was done using the facilities
of the Supercomputer Center,
Institute for Solid State Physics, University of Tokyo.

\end{document}